\documentstyle[preprint,prl,aps]{revtex}
\begin{document}
\tightenlines
\draft
\title{Bi-stability of mixed states in neural network \\ storing
hierarchical patterns
}
\author{Kaname TOYA}
\address{
Department of Systems and Human Science, Graduate School of
Engineering Science, Osaka University, 1-3, Machikaneyama, Toyonaka,
Osaka 560-8531, Japan
}

\author{Kunihiko FUKUSHIMA}
\address{Department of Information and Communication Engineering, 
The University of Electro-Communications,
1-5-1, Chofugaoka, Chofu, Tokyo 182-8585, Japan
}

\author{Yoshiyuki Kabashima}
\address{Department of Computational Intelligence and Systems Science,
Interdisciplinary Graduate School of Science and Engineering, 
Tokyo Institute of Technology, 
Yokohama 226-8502, Japan}

\author{Masato OKADA}
\address{Kawato Dynamic Brain Project, Japan Science and Technology
Corporation, 2-2, Hikaridai, Seika-cho, Soraku-gun, Kyoto 619-0288,
Japan
}

\date{\today}
\maketitle

\begin{abstract}
We discuss the properties of equilibrium states in an autoassociative
memory model storing hierarchically correlated patterns (hereafter,
hierarchical patterns). We will show that symmetric mixed states
(hereafter, mixed states) are bi-stable on the associative
memory model storing the hierarchical patterns in a region of the ferromagnetic
phase. This means that the first-order transition occurs in this
ferromagnetic phase. We treat these contents with a statistical
mechanical method (SCSNA) and by computer simulation. Finally, we
 discuss a physiological implication of this model.
Sugase et al.\ analyzed the time-course of the information carried by the firing
of face-responsive neurons in the inferior temporal cortex \cite{Sugase1999}. We also
discuss the relation between the theoretical results and the
physiological experiments of Sugase et al.\
\end{abstract}
\pacs{PACS numbers: 84.35.+i, 87.10.+e, 05.90.+m, 89.70.+c}

\narrowtext

\section{Introduction}

There are two kinds of hypotheses regarding internal representations of memory items in the
brain. First, there is the ``distributed representation'' hypothesis, which assumes that our
memory items are encoded by neuron activity patterns. But another hypothesis
(e.g., the ``grand  mother cell'' hypothesis) has been proposed,
where the memory items are represented by the excitation of
corresponding neurons: that is, the ``local representation''
hypothesis. Let us consider a neural network
model consisting of $N$ neurons. The number of memory items in the local
representation is $O(N)$.
One might think that the number of memory
items in the distributed representation is $O( 2^N)$ 
in the case of storing binary patterns.
However, this consideration
is meaningless. If we consider error correcting abilities, this becomes $O(N)$
on the basis of the statistical mechanical theories (for example
\cite{AGS1985}). One of the remarkable
advantages of the distributed representation is that the relationship between the
memory items is naturally implemented by the distance of memory patterns.
However, many studies on associative memory models have been
confined to those with uniformly distributed patterns. Thus, we should discuss a
model that stores some structural patterns. We will extend the conventional
SCSNA \cite{Shino1992} to a generalized SCSNA in order to treat a model with any structural
pattern and a general class of output functions of neuron.

When an associative memory model is made to store memory patterns as a
result of correlation learning, a pattern, which has an uniform overlap 
with some of the stored memory patterns, automatically becomes the equilibrium state of the model.
This is called the mixed state. It is not appropriate to think that
this mixed state is a side-effect and/or that it is unnecessary for
information processing. Amari has discussed a ``concept formation''
using the stability of mixed state \cite{Amari1977}. The correlated attractor
\cite{Amit1993,Amit1994} accounting for
the physiological experiments by Miyashita \cite{Miyahsita1998} can be interpreted as a 
mixed state in a broad sense. Recently, Parga and Rolls \cite{Parga} used the mixed state 
in their research on the mechanism of invariant recognition with a 
coordinate transformation in the visual system.

The aim of this research is to study the properties of mixed states in
the autoassociative memory model storing hierarchical patterns by
utilizing the generalized SCSNA and computer simulation. First, we discuss a
model that stores patterns in which a two-stage hierarchy exists. With the generalized SCSNA, we derive the order parameter equations for
the equilibrium state in this model. By solving the obtained
equations, we show that two kinds of mixed states coexist in a particular
region of the retrieval phase (ferromagnetic phase). It is a
characteristic of the two kinds of mixed states that they have different
values of cross-talk noise variances, that is, they are influenced by the
uncondensed patterns in different ways. This kind of
bi-stability in the retrieval phase has not been previously reported.
We will show that the bi-stability of mixed states does not depend on the
number of patterns in the same cluster.
Generally speaking, it is not so easy to confirm the multi-stable states by computer simulation.
However, we succeed in confirming the bi-stability of the mixed
states by computer simulation after considering the qualitative properties of the
retrieval dynamics in this model.
Next, we treat a model storing a set of hierarchical patterns and
uniformly distributed patterns in order to investigate universality of the
bi-stability. From these theoretical results, we found
that a bi-stability of mixed states also exists in this model.
Contrary to these models,
such bi-stability of mixed states does not exist in a model where the
contribution of uncondensed patterns to the synaptic couplings is
replaced by the spin glass type interaction. We will discuss the
reason for this by using the SCSNA.

Recently, Sugase et al.\ have reported interesting phenomena
concerning the temporal dynamics of face-responsive neurons in the
inferior temporal (IT) cortex \cite{Sugase1999}. In the discussion, we will examine the relation
between the obtained results and the
physiological experiments by Sugase et al.
\section{Model}
Let us consider a recurrent neural network consisting of $N$ neurons
with an output function $F(\cdot)$. We employ the synchronous dynamics,
 \begin{eqnarray}
x_i^{t+1} &=& F ( \sum_{j \neq i}^N J_{ij} x_j^{t} ), \label{dynamics.rule}
\end{eqnarray}
where $x_i^{t}$ represents a state of the $i$-th neuron at discrete time 
$t$, and discuss the case of the thermodynamics limit ($N \rightarrow \infty $).
$J_{ij}$ in the above equation denotes a synaptic coupling from the $j$-th neuron to the
$i$-th neuron.
In this work, we discuss the equilibrium state of eq. (\ref{dynamics.rule}). 
For simplicity, we treat a two-stage hierarchy, which is one of the
simplest cases.
This can be
easily extended to more complex hierarchies.
One can use many procedures to make the set of ultrametric patterns, but 
we employ the following method. Each component $\xi_i^{\mu,  \nu}$ of
pattern $\mbox{\boldmath$\xi$}^{\mu,  \nu}$ is a
random variable drawn from the following probability distributions,
\begin{eqnarray}
\mbox{P}[\xi_i^{\mu } = \pm 1]& =& \frac{1}{2},\ \ \   i=1,  2,   \cdots,
 N \, \ \mu=1,  2,   \cdots, p,  \label{make.parent} \\
\mbox{P}[\xi_i^{\mu, \nu} = \pm 1]& =& \frac{1 \pm b \xi^\mu_i}{2},\ \
 \nu=1,  2,  \cdots,  s 
\label{make.children}
\end{eqnarray}
with $0 \le b \le 1$.
The distance between patterns $\mbox{\boldmath$\xi$}^{\mu,  \nu}$ is expressed by
\begin{eqnarray}
 \mbox{E}[\xi^{\mu, \nu}_i \xi^{\mu', \nu'}_i] &=&  \delta_{\mu \mu'} (\mbox{\bf B})_{\nu \nu'},    
  \label{overlap}\\
  (\mbox{\bf B})_{\nu \nu'}&\equiv&\delta_{\mu \mu'} (\delta_{\nu
   \nu'}+b^2(1-\delta_{\nu \nu'}) ) \label{Bdefinition},
\end{eqnarray}
where $\mbox{E}[\cdot]$ stands for an average with respect to the
probabilities distributed in eqs.\ (\ref{make.parent}) and
(\ref{make.children}), and $\delta_{\mu \mu'}$ is the Kronecker's
$\delta$ defined as,
\begin{eqnarray}
\delta_{\mu \mu'}&=&\left\{
\begin{array}{ll}
1 & (\mu = \mu' ) \\
0 & (\mu \neq \mu')
\end{array} 
\right. .
\end{eqnarray}
According to the definition of the matrix {\bf B} in eq.\ (\ref{Bdefinition}), {\bf B} is $s\times s$ matrix.
As shown in eq.\ (\ref{overlap}), the memory patterns
$\mbox{\boldmath$\xi$}^{\mu,  \nu}$ have a two-stage ultrametric
structure. $(\mbox{\bf B})_{\nu \nu'}$ in eq.\ (\ref{overlap}) stands
for the element in the $\nu$-th row, the $\nu'$-th column of matrix {\bf
B}. $\mbox{\boldmath$\xi$}^{\mu,  \nu}$ are $ps$ uniformly
generated patterns when $b=0$, while $\mbox{\boldmath$\xi$}^{\mu,  \nu}$
in the $\mu$-th cluster are the same when $b=1$.

We employ the simple Hebbian rule as the learning rule, and the synaptic coupling $J_{ij}$ is set to
\begin{eqnarray}
J_{ij}& =& \frac{1}{N}\sum_{\mu=1}^{\alpha N} \sum_{\nu=1}^{s}
	\xi^{\mu, \nu}_i \xi^{\mu, \nu}_j 
\label{equation.learning_model1}
\end{eqnarray} 			
where $\alpha=p/N$. Since the number of clusters is $\alpha N$, we call $\alpha$ the
loading rate. $\mbox{\boldmath$\xi$}^{\mu}$ is not explicitly
used for the learning rule. We discuss two kinds of models. 
One is the previously defined model with the couplings $J_{ij}$ given
by eq.\ (\ref{equation.learning_model1}), which we call Model 1.
The other model, Model 2, is shown here with the following couplings $J_{ij}$,
\begin{eqnarray}
J_{ij}& =& \frac{1}{N}\sum_{\nu=1}^{s}
	\xi^{1, \nu}_i \xi^{1, \nu}_j +\frac{1}{N}
\sum_{\mu=2}^{\alpha N} \xi^{\mu}_i \xi^{\mu }_j. 
\label{equation.learning_2}
\end{eqnarray}    
We will examine the stability of the memory pattern and a mixed state
which has a uniform overlap with an odd number of memory patterns in
the same cluster. There exist mixed states which have finite overlaps
with memory patterns in different clusters. However, we do {\it not}
discuss this topic in the present paper.
\section{A generalized SCSNA}
There are two kinds of statistical mechanical theories treating the
equilibrium properties of associative memory models. One is based on 
the SCSNA proposed by Shiino and Fukai \cite{Shino1992}. The other one
is the replica theory, which has been used to analyze the equilibrium
states in Model 1 \cite{Fontanari1990}. However, the replica
theory cannot be applied to a system where the free energy
cannot be defined (e.g., a model with a nonmonotonic output
function \cite{ShinoFukai1993} or an oscillator associative memory
model \cite{Aonishi}), while the SCSNA can treat these systems without
the energy function. Since previous studies
with the SCSNA have mainly focused on systems with uniformly
distributed patterns, the SCSNA cannot be directly applied to Model 1.

We will extend the previously proposed SCSNA to
a generalized SCSNA in this section
in order to treat a model storing a set of structural patterns $\mbox{\boldmath
$\sigma$}^{\rho} \ (\rho=1,  2, ..., \hat{\alpha} N)$ randomly generated 
by an independent identical probability distribution with respect to $i$. 
The correlation matrix $(\mbox{\bf C})_{\rho \rho'}$ between the two
patterns $\mbox{\boldmath $\sigma$}^{\rho}$ and $\mbox{\boldmath
$\sigma$}^{\rho'}$ is given by,
\begin{eqnarray} 
 (\mbox{\bf C})_{\rho \rho'}&\equiv&
\frac{1}{N} \sum_{i=1}^{\hat{\alpha} N} \sigma^{\rho}_i \sigma^{\rho'}_i 
,\\
&=& \mbox{E}[\sigma^{\rho}_i
  \sigma^{\rho'}_i]. \label{equation.general_overlap}
\end{eqnarray}
We consider a recurrent network
consisting of $N$ neurons with the output function $F(\cdot)$ and
synaptic couplings $J_{ij}$ given by
\begin{eqnarray}
J_{ij}& =& \frac{1}{N} \sum_{\rho=1}^{\hat{\alpha} N} \sigma^{\rho }_i
 \sigma^{\rho }_j. \label{gene_hebb} 
\end{eqnarray}
Since the number of memory patterns is $\hat{\alpha} N$, $\hat{\alpha}$ is
defined as the loading rate in this model.
$\mbox{\boldmath $\xi$}^{\mu, \nu}$ or ${\bf B}$ in eq.\ (\ref{overlap})
is an example of $\mbox{\boldmath $\sigma$}^{\rho}$ or {\bf C} in
eq.\ (\ref{equation.general_overlap}), respectively, 
\begin{eqnarray}
\mbox{\boldmath $\sigma$}^{\rho}&\leftrightarrow&\mbox{\boldmath
 $\xi$}^{\mu,  \nu} \ \ \  \rho=s(\mu-1)+\nu,  \label{xisigma_relation}\\
({\bf C})_{\rho \rho'}&\leftrightarrow&\delta_{\mu \mu'} ({\bf
 B})_{\nu \nu'}, \label{BC_relation}\\
\hat{\alpha} &\leftrightarrow& \alpha s. \label{alpha_relation}  
\end{eqnarray}
We consider the case where the state of neurons at discrete time step
$t$, $\mbox{\boldmath $x$}^t$, is synchronously updated, 
\begin{eqnarray}
x_i^{t+1} &=& F ( \sum_{j \neq i}^N J_{ij} x_j^{t} ), \label{dynamics_gene.rule}
\end{eqnarray}
and discuss the equilibrium state
$\mbox{\boldmath $x$}$ with the limit $t \rightarrow \infty $.
We introduce a set of rotated memory patterns, $\mbox{\boldmath
$\bar{\sigma}$}=\{\bar{\sigma_i}^1,  \bar{\sigma_i}^2,
. . ., \bar{\sigma_i}^{\hat{\alpha} N }\}$, as
\begin{eqnarray}
 \bar{\sigma}^{\rho }_i &=& \frac{1}{\sqrt{\kappa_\rho}}
  \sum_{\rho'  =1}^{\hat{\alpha} N} W_{\rho \rho'} \sigma^{\rho'}_i,   \label{equation. gtrans}\\
 \mbox{\bf W} &=& (\mbox{\boldmath $w$}^1,   \mbox{\boldmath $w$}^2,   \cdots,  
  \mbox{\boldmath $w$}^{\hat{\alpha} N})^{\mbox{T}}. 
\end{eqnarray}
$\mbox{\boldmath$w$}^{\rho}(\rho=1,  2,  . .., \hat{\alpha}N)$
and $\kappa_{\rho}(\rho=1,  2, . . .,\hat{\alpha}N)$ in the above equations represent 
the $\rho$-th $\hat{\alpha}N$ dimensional normalized eigenvector of matrix
$\mbox{\bf C}$ and the $\rho$-th eigenvalue of matrix $\mbox{\bf C}$ for
$\mbox{\boldmath$w$}^{\rho}$, respectively.
Each component $\bar{\sigma}^{\rho }_i$ is statistically
dependent (independent of) on $\mu$ ($i$), respectively, and
satisfies the following conditions,
\begin{eqnarray}
\frac{1}{N} \sum_{i=1}^N  (\bar{\sigma}^{\rho }_i)^2 &=& 1 ,\\
\frac{1}{N} \sum_{i=1}^N  \bar{\sigma}^{\rho }_i \bar{\sigma}^{\rho' }_i
 & \sim &O(\frac{1}{\sqrt{N}}),\ \ \rho \neq \rho'.  
\end{eqnarray}
Using the rotated patterns, we can rewrite $J_{ij}$ in eq.\ (\ref{gene_hebb}) as
\begin{eqnarray}
J_{ij}&=& \frac{1}{N} \sum_{\rho=1}^{\hat{\alpha} N} \kappa_{\rho}
 \bar{\sigma}^{\rho }_i \bar{\sigma}^{\rho }_j. 
\end{eqnarray}
The overlaps $\bar{m}^{\rho}$ between the equilibrium state $\mbox{\boldmath $x$}$
and $\bar{\mbox{\boldmath $\sigma$}}^{\rho}$ are defined by the
following equation, 
\begin{equation}
\bar{m}^{\rho} = \frac{1}{N}\sum_{i=1}^N
\bar{\sigma}_i^{\rho} x_i .\label{overlap_before_sigma} 
\end{equation} 
If one assumes that the equilibrium state $ \mbox{\boldmath
$x$}$ has nonzero overlaps with $\bar{s}$ rotated patterns $ \bar{\mbox{\boldmath $\sigma$}}^{\rho}(1 \le
\rho \le \bar{s})$, we can derive the SCSNA order parameter equations
\cite{Mimura1998} (see Appendix A), 
\begin{eqnarray}
\bar{m}^{\rho}&=& \int Dz < \bar{\sigma}^{\rho} Y(z
 ,  \bar{\sigma}^{1},  \bar{\sigma}^{2},  . . ,  \bar{\sigma}^{\bar{s}}) >_{\bar{\mbox{\boldmath $\sigma$}}},  
\label{equation.gm} \\
q&=& \int Dz <( Y(z,  \bar{\sigma}^{1},  \bar{\sigma}^{2},  . . ,  \bar{\sigma}^{\bar{s}}))^2 >_{\bar{\mbox{\boldmath $\sigma$}}},  
\label{equation.gq}\\
U &=& \frac{1}{\sqrt{\hat{\alpha}r}}\int Dzz < Y(z,  \bar{\sigma}^{1},  \bar{\sigma}^{2},  . . ,  \bar{\sigma}^{\bar{s}})
 >_{\bar{\mbox{\boldmath $\sigma$}}} ,  
\label{equation.gU}\\
Dz &=& \frac{dz}{\sqrt{2\pi}}\exp(\frac{-z^2}{2}) ,\\
Y(z,  \bar{\sigma}^{1},  \bar{\sigma}^{2},  . . ,  \bar{\sigma}^{\bar{s}}) &=&F(\sum_{\rho=1}^{\bar{s}}
 \kappa_{\rho} \bar{\sigma}^{\rho}\bar{m}^{\rho} +\Gamma Y(z,  \bar{\sigma}^{1},  \bar{\sigma}^{2},  . . ,  \bar{\sigma}^{\bar{s}}) +\sqrt{\hat{\alpha} r}z),   \label{equation.gY}\\
r &=& q \int_0^1 du \frac{ \kappa (u)^{2}  }{ ( {1 -
 \kappa (u) U)}^2} =  \frac{q}{\hat{\alpha} N} {\bf Tr} \biggl( \frac{{\bf
 C }^{2} }{ ( {\bf I - C}U)^2} \biggl) ,  \label{equation.gr}\\
\Gamma &=& \hat{\alpha}  \int_0^1 du \frac{ {\kappa (u)}^{2} U }{ 1 -
 \kappa (u) U} =  \frac{1}{ N} {\bf Tr} \biggl( \frac{{\bf
 C }^{2}U }{  {\bf I - C}U }\biggl) . \label{equation.gGamma}
\end{eqnarray} 
where $<\cdots>_{\mbox{\boldmath $\bar{\sigma}$}}$ stands for an average
over the condensed patterns $\mbox{\boldmath $\bar{\sigma}$}=(\bar{\sigma}^1,  \bar{\sigma}^2
,  . . . ,  \bar{\sigma}^{\bar{s}})$.
We can express the sum of $\kappa_{\rho}$
in terms of an integration along continuous eigenvalue $\kappa
( \frac{\rho}{\hat{\alpha} N} ) \equiv \kappa_{\rho}$ for $p, N
\rightarrow \infty$.
We remarked that the analytical expressions of $r$ and $\Gamma$ given by 
eqs.\ (\ref{equation.gr}) and (\ref{equation.gGamma}) depend only
on the matrix {\bf C} and do not explicitly depend on the condensed patterns
$\mbox{\boldmath $\bar{\sigma}$}^{\rho} (\rho=1,2,...,\bar{s})$.
This fact leads to the following order parameter equations for the
equilibrium state having nonzero overlaps $ \displaystyle m^{\rho}=\frac{1}{N} \sum_{i=1}^N
\sigma_i^{\rho} x_i $ with $s$ original memory patterns $\mbox{\boldmath
$\sigma^{\rho}$} \ (\rho=1,2,...,s)$:
\begin{eqnarray}
m^{\rho}&=&\int Dz <\sigma^{\rho} Y(z,  \sigma^{1},  \sigma^{2},  . . ,  \sigma^{s}) >_{\mbox{\boldmath $\sigma$}}, \label{equation.xigm}\\
q&=& \int Dz < (Y(z
 ,  \sigma^{1},  \sigma^{2},  . . , \sigma^{s}))^2 >_{\mbox{\boldmath $\sigma$}},  
\label{equation.xigq}\\
U &=& \frac{1}{\sqrt{\hat{\alpha} r}} \int Dzz <Y(z
 , \sigma^{1},  \sigma^{2},  . . ,  \sigma^{s}) >_{\mbox{\boldmath $\sigma$}},  
\label{equation.xiU}\\
Dz &=& \frac{dz}{\sqrt{2\pi}}\exp(\frac{-z^2}{2}) ,\\
Y(z,  \sigma^{1},  \sigma^{2},  . . ,  \sigma^{s}) &=&F(\sum_{\rho=1}^{s}
 \sigma^{\rho} m^{\rho} +\Gamma Y(z,  \sigma^{1},  \sigma^{2},  . . ,  \sigma^{s}) +\sqrt{\hat{\alpha} r}z),   \label{equation.xigY}\\
r &=&  \frac{q}{\hat{\alpha} N} {\bf Tr} \biggl( \frac{{\bf
 C }^{2} }{ ( {\bf I - C}U)^2} \biggl) ,  \label{equation.xigr}\\
\Gamma &=& \frac{1}{ N} {\bf Tr} \biggl( \frac{{\bf
 C }^{2}U }{ {\bf I - C}U} \biggl) . \label{equation.xigGamma}
\end{eqnarray}
Note that the off-diagonal terms of the matrix {\bf C} between the condensed and uncondensed
pattern spaces can be neglected if $\bar{s}$ is taken to be sufficiently
large but $O(1)$.
\section{Results}
We discuss the case where $F(\cdot)= \mbox{sgn}(\cdot)$ defined by
\begin{eqnarray}
\mbox{sgn}(u)&=&\left\{
\begin{array}{ll}
1 & (u \ge 0  ) \\
-1& (u < 0 )
\end{array} 
\right. .
\end{eqnarray}
First, we will show the results of the analysis for the equilibrium properties in
Model 1.
Now, let us apply the generalized SCSNA proposed in section III to
Model 1.
We rewrited eq.\ (\ref{BC_relation}) by
\begin{equation}
{\bf C}= \underbrace{\bf{B} \bigoplus \bf{B} \bigoplus,  . . . ,  \bigoplus \bf{B}}_{p}.\label{BC_relation2}
\end{equation}
The matrix {\bf B} has the eigenvalues  $\lambda_{1}=1+(s-1) b^2 ,
\lambda_\nu=1-b^2 (2 \le \nu \le s)$. One assumes that the overlaps
$ \displaystyle m^{1,  \nu}=\frac{1}{N} \sum_{i=1}^N \xi^{1,\nu}_i x_i \
\ (\nu=1,2,...,  s)$ have the values of $O(1)$.
By considering the relations given by eqs.\ (\ref{xisigma_relation}),
(\ref{alpha_relation}) and (\ref{BC_relation2}), we obtained
the following SCSNA order parameter equations for this model:
\begin{eqnarray}
m^{1,  \nu}&=& < \xi^{1,  \nu}\mbox{erf}
(  \frac{ \sum_{\sigma = 1}^{s} \xi^{1, \sigma}m^{1,\sigma} }{ \sqrt{2 \alpha r} } )
 >_{{\mbox{\boldmath $\xi$}}^1}, \label{equation.model1m}   \\
q &=&1, \label{equation.model1q} \\
r &=& q \sum_{\nu=1}^s \biggl( \frac{{ \lambda_{\nu} }^{2} }{ ( 1 -
 \lambda_{\nu}U)^2} \biggl), \label{equation.model1r} \\
U &=& \sqrt{ \frac{2}{\pi \alpha r} }
< \exp (
- (\sum_{ \sigma=1}^{s}\frac{  \xi^{1, \sigma} m^{1, \sigma }}{\sqrt{2 \alpha r}})^2 ) 
 >_{{\mbox{\boldmath $\xi$}}^1}. \label{equation.model1U} 
\end{eqnarray} 
where $< \cdots >$ stands for an average over the condensed patterns
${\mbox{\boldmath $\xi$}}^1=( \xi^{1,1},\xi^{1,2},\cdots,\xi^{1,s}) $.
In the above equations, we omitted the $\Gamma$ terms corresponding to
eq.\ (\ref{equation.xigGamma}) by applying the Maxwell rule.
The results coincided with those of the theoretical
analysis using the replica theory \cite{Fontanari1990}. 
For simplicity, we will show the case where
$s=3,5$. Fig.\ \ref{phasebalpha} shows a phase diagram of numerical
solutions for
eqs.\ (\ref{equation.model1m}) - (\ref{equation.model1U}). The critical
loading rate of a memory pattern and that of a mixed state are plotted against
$b$. The triangular area shown in the figure represents the region where two kinds of mixed
states coexist. We called this region the ``bi-stable region''. In this
paper, a mixed pattern which is similar to $\mbox{sgn}(\sum_{\nu=1}^s
\mbox{\boldmath $\xi$}^{\mu,\nu})$, is defined as $\mbox{\boldmath $\eta$}^{\mu}$,
and the other one, which is reported here for the first time, is defined as
$\mbox{\boldmath $\tilde{ \eta}$}^{\mu}$. As shown in \cite{Shino1992}, 
the SCSNA assumes the stability of equilibrium state. There is no
free energy in the SCSNA formalism. However, we discuss the present
model using the equilibrium statistical mechanics \cite{Fontanari1990}
since it has the
energy function. From the statistical-mechanical
viewpoint, the first-order transition occurs at the dashed curve in
Fig.\ \ref{phasebalpha}. In the region below the dashed curve, the free
energy of $\mbox{\boldmath$\eta$}^{\mu}$ is less than that of
$\tilde{\mbox{\boldmath$\eta$}}^{\mu}$,
while the free energy of $\mbox{\boldmath$\eta$}^{\mu}$ is larger than
that of $\tilde{\mbox{\boldmath$\eta$}}^{\mu}$ in the region above
the dashed curve.

We will examine two typical examples in the bi-stable
region. We consider the case where the retrieval pattern is $\mbox{\boldmath
$\xi$}^{1,1}$. $\alpha^{\eta}_c$ ($ \alpha^{\tilde{\eta}}_c$) is defined as the
critical loading rate of $\mbox{\boldmath $\eta$}^{1}$ ( $\mbox{\boldmath
$\tilde{ \eta}$}^{1}$ ). First, we discuss the region in which $\alpha^{\eta}_c
<\alpha^{\tilde{\eta}}_c$ as follows. Fig.\ \ref{b61malpha}(a) shows how
the overlaps between the equilibrium state and the retrieval pattern
$\mbox{\boldmath $\xi$}^{1,1}$ depend on $\alpha$.
Two kinds of mixed states coexist when $0. 01500 < \alpha < 0.01765$, as
shown in Fig.\ \ref{b61malpha}(a). From the numerical analysis, the
value of the cross-talk
noise variance expressed by $r$ in eq.\ (\ref{equation.model1r})
becomes larger in the order $\mbox{\boldmath $\xi$}^{1,1}$, $\mbox{\boldmath
$\eta$}^{1}$, $\mbox{\boldmath $\tilde{ \eta}$}^{1}$.
The variation of the overlap $\displaystyle \frac{1}{N} \sum_{i=1}^{N}
\xi^{1,1}_i \tilde{ \eta}_i^{1}$ with $\alpha$ is larger than that of 
$\displaystyle \frac{1}{N} \sum_{i=1}^{N} \xi^{1,1}_i
\eta_i^{1}$ with $\alpha$, as shown in Fig.\ \ref{b61malpha}(a).
Thus, we found that $\mbox{\boldmath $\tilde{ \eta}$}^{1}$ is more influenced by
the uncondensed patterns than $\mbox{\boldmath $\eta$}^1$ is.
We will transform the the order parameter equations given by
eqs.\ (\ref{equation.model1m}) - (\ref{equation.model1U})
into an equation in one
variable in order to qualitatively understand the nature of this
bi-stability.
Considering that $ m^{1,\nu} $ is the same value for any $\nu$ in the mixed
states, we can transform the order parameter equations representing
the mixed states into the equation
in one variable $y$ by replacing $\frac{m^{1, \nu}}{\sqrt{2 \alpha r}}$
with $y^{\nu}$ (see Appendix B).
Fig.\ \ref{b61malpha}(b) shows the graphical solutions of the transformed equation around the
bi-stable region. Each intersection of the $y$-axis and function
$\Phi(y,  \alpha, b=0. 61,  s=3)$ represents the corresponding solution.
Fig.\ \ref{b61malpha}(b) shows that two kinds of mixed states, that is,
$\mbox{\boldmath $\eta$}^{1}$ and $\mbox{\boldmath $\tilde{ \eta}$}^{1}$,
coexist at $\alpha=0. 016$. On the other hand,
there is no such bi-stability in the retrieval region except for the
triangular region. Fig.\ 1 shows that bi-stability also exists for $s=5$.
By analyzing the graphical solutions of the transformed equation for
various values of $s$, 
we deduced that the bi-stability of
the mixed states exists for any value of $s$ in Model 1. 

We performed computer simulation to confirm the bi-stability of mixed states, and
obtained equilibrium states, which are expressed
by the dotted curve (A) in Fig.\ \ref{b61malpha}(a), as follows.
We first got an equilibrium state at $\alpha=0.0003 < \alpha^{\eta}_c$ with the
initial state set to $\mbox{sgn}(\sum_{\nu=1}^s \mbox{\boldmath $\xi$}^{1,\nu})$.
After that, equilibrium states for various $\alpha$ were obtained by
gradually increasing the value of $\alpha$ from $\alpha=0.0003$. Since
the simulation results mostly agreed with the theoretical results, we could not
distinguish between them as shown in this figure.
The dotted curve (B) in Fig.\ \ref{b61malpha}(a) was obtained as follows. An
equilibrium state at $\alpha=0.019$ ($\alpha^{\eta}_c=0.01765 < \alpha < 
\alpha^{\tilde{\eta}}_c=0.01982$ ) was obtained by gradually increasing the value of
$\alpha$ from $\alpha=0.0003$ in computer simulation.
Using this equilibrium state, we got each equilibrium state
at the corresponding $\alpha$ by gradually decreasing the value of $\alpha$ from 
$\alpha=0.019$.
We carried out computer simulation with $N=10000$; 
a typical example is shown in Fig.\ \ref{b61malpha} (a). 
Fig.\ \ref{b55malpha}(a) shows a typical example ($b=0. 55,  s=3$) of the
case where $\alpha^{\tilde{\eta}}_c < \alpha^{\eta}_c$. 
Two kinds of mixed states coexisted when $0. 01164 < \alpha < 0. 01389$.
The axes in this figure are the same axes in
Fig.\ \ref{b61malpha}(a). Fig.\ \ref{b55malpha}(b) shows the graphical solutions of the
transformed equation in one variable around the bi-stable region
when $s=3,  b=0. 55$. This figure shows that two kinds of mixed states,
that is, $\mbox{\boldmath $\tilde{ \eta}$}^{1}$ and
$\mbox{\boldmath$\eta$}^{1}$, are stable at $\alpha=0.0125$.
The dotted curve (C) in Fig.\ \ref{b55malpha}(a) was obtained as follows.
We first got an equilibrium state at $\alpha=0.0003$ with the
initial state set to $\mbox{sgn}(\sum_{\nu=1}^s \mbox{\boldmath $\xi$}^{1,\nu})$.
After that, equilibrium states for various $\alpha$ were obtained by
gradually increasing the value of $\alpha$ from $\alpha=0.0003$.
Since $\mbox{ \boldmath ${\eta}$}^1$ is always stable when
$\alpha^{\tilde{\eta}}_c < \alpha < \alpha^{\eta}_c$, an equilibrium state
corresponding to $\mbox{ \boldmath $\tilde{\eta}$}^1$ could not be
obtained by computer simulation with the initial state set to $\mbox{sgn}(\sum_{\nu=1}^s \mbox{\boldmath
$\xi$}^{1,\nu})$. By considering the qualitative properties of the retrieval
dynamics, we carried out computer simulation in the following
manner. Fig.\ \ref{dynamics}
shows the trajectories of temporal evolutions of overlaps of the
state $\mbox{\boldmath $x$}^t$ with $\mbox{\boldmath $\xi$}^{1,1}$ or
$\mbox{\boldmath $\xi$}^{1,2}$ respectively,
\begin{eqnarray}
\displaystyle m^{1,\nu}_{t} &=& \frac{1}{N} \sum_{i=1}^N \xi^{1,\nu} x^t_i.
\end{eqnarray}
The parameters were set to $s=3, b=0. 475, \alpha=0. 0087, N=40000$.
The initial states were set as $\displaystyle \mbox{P}[x_i^{0} = \pm 1] = \frac{1 \pm m^{1,1}_0
\xi^{1,  1}_i}{2}$ with an initial overlap $ m^{1,
1}_{0}$. Theoretically, $m^{1,2}_t = m^{1,3}_t$ holds over the retrieval process in this case.
crosses that are shown in Fig.\ \ref{dynamics} represent the
theoretically obtained equilibrium states corresponding to $\mbox{\boldmath
$\tilde{ \eta}$}^1$, $\mbox{\boldmath $\eta$}^1$ and
$\mbox{\boldmath $\xi$}^{1,  1}$.
Two kinds of mixed states ($\mbox{\boldmath $\eta$}^1$, $
\mbox{\boldmath $\tilde{\eta}$}^1$ ) and the retrieval pattern ($\mbox{\boldmath
$\xi$}^{1,  1}$) were stable at this parameter set (\ \ Fig.\ \ref{dynamics}).
The dynamics converged to
$\mbox{\boldmath $\tilde{ \eta}$}^1$, $\mbox{\boldmath $\eta$}^1$,
$\mbox{\boldmath $\xi$}^{1,  1}$ in order as $ m^{1,  1}_{0}$
increased, while the cross-talk noise variance for the equilibrium became larger in the order,
$\mbox{\boldmath $\xi$}^{1,  1}, \mbox{\boldmath $\eta$}^{1},
\tilde{\mbox{\boldmath $\eta$}}^1$. This phenomenon occurred because the
cross-talk noise variance of the initial
stage of the retrieval process increased as $m^{1,1}_{0}$ decreased.
The results showed that equilibrium states
corresponding to $\mbox{\boldmath $\tilde{ \eta}$}^1$ can be confirmed
by computer simulation when the initial state
has an appropriate overlap with the retrieval pattern.
The dotted curve (D) in Fig.\ \ref{b55malpha}(a) was obtained as follows. We got an
equilibrium state at $\alpha=0.0135$ where the initial state had the
appropriate overlap with the retrieval pattern ($\mbox{\boldmath $\xi$}^{1,  1}$).
We obtained equilibrium states for various $\alpha$
by increasing (decreasing) the value of $\alpha$ from $\alpha=0.0135$.
We carried out computer simulation with $N=10000$; Fig.\ 3(a) shows a
typical example. Although the result from this simulation did not
always quantitatively agree with the theory owing to spurious states,
we can state that the bi-stability of two kinds of mixed states exists
in the particular region of the retrieval phase from the hysteresis
shown in Figs.\ \ref{b61malpha}(a) and \ref{b55malpha}(a).

Let us present the equilibrium properties in Model 2. Since the uncondensed
patterns $\mbox{\boldmath $\xi$}^{\mu} (\mu=2,3,...,\alpha N)$ of
Model 2 satisfy the orthogonal condition ${\mbox E}[\xi^{\mu}_i
\xi_i^{\mu'}]=\delta_{\mu \mu'}$, the order
parameter equations for this model were derived by replacing
eq.\ (\ref{equation.model1r}) with eq.\ (\ref{equation.model2r}):
\begin{eqnarray}
m^{1,  \nu}&=& < \xi^{1,\nu}\mbox{erf}
(  \frac{ \sum_{\sigma = 1}^{s} \xi^{1, \sigma}m^{1, \sigma} }{ \sqrt{2 \alpha r} } )
 >_{{\mbox{\boldmath $\xi$}}^1}, \label{equation. model2m}   \\
q &=&1,\label{equation. model2q} \\
r &=& \frac{q }{ ( 1 -U)^2 }  ,\label{equation.model2r} \\
U &=& \sqrt{ \frac{2}{\pi \alpha r} }
< \exp (
-(\sum_{ \sigma=1}^{s}\frac{\xi^{1, \sigma} m^{1, \sigma } }{\sqrt{2
\alpha r } })^2 ) 
 >_{{\mbox{\boldmath $\xi$}}^1}. \label{equation. model2U} 
\end{eqnarray} 
We transformed the order parameter equations concerning the
mixed states into an equation
in one variable $y^{\nu}=\frac{m^{1,\nu}}{\sqrt{2\alpha r}}$ (see Appendix B).
As in Model 1, we deduced
that there is bi-stability of mixed states for an arbitrary $s$.
These results indicated that this bi-stability for mixed states is not
dependent on details of the structure in stored patterns.

\section{Discussion}
We have shown that bi-stability of mixed states exists in two typical
models, Model 1 and Model 2. The reason for this
bi-stability, which is based on the SCSNA, can be explained as follows.
We introduce a model where the
contribution of uncondensed patterns to the coupling $J_{ij}$ is
replaced by the spin glass type interaction, and we call this model
Model 3. 
The synaptic coupling of Model 3 is defined as follows,
\begin{equation}
J_{ij} =\frac{1}{N} \sum_{\nu=1}^s \xi^{1,\nu}_i\xi^{1,\nu}_j+ \delta_{ij},
\end{equation}
and the symmetric noise $\delta_{ij}$ is independently drawn from the following rule,
\begin{eqnarray}
\delta_{ji} &\sim &N(0, \frac{\delta^2}{N}) ,\\
\delta_{ij}&=&\delta_{ji} ,
\end{eqnarray}
where $\delta$ is constant.
We obtained the following order parameter equations for this model with
the SCSNA\cite{Okadaetal1998},
\begin{eqnarray}
m^{1,\nu}&=& \int Dz < \xi^{1,\nu}Y >_{{\mbox{\boldmath $\xi$}}^1},  
\label{equation.Model3m} \\
q&=& \int Dz < (Y)^2 >_{{\mbox{\boldmath $\xi$}}^1}, 
\label{equation.Model3q}\\
U&=& \frac{1}{\sigma} \int Dzz < Y>_{{\mbox{\boldmath $\xi$}}^1}, 
\label{equation.Model3U}\\
Y &=& F(\sum_{\sigma=1}^s \xi^{1,\sigma} m^{1,\sigma}+ \Gamma Y +\sigma z),\\
\Gamma &=& {\delta}^2 U ,\label{equation.Model3gamma}\\
Dz &=& \frac{dz}{\sqrt{2\pi}}\exp(\frac{-z^2}{2}) ,\\
{\sigma}^2 &=& {\delta}^2 q .\label{equation.Model3r}
\end{eqnarray} 
As you can notice, this model corresponds to the Sherrington-Kirkpatrick (SK) model.
In contrast to Model 1 or Model 2, a bi-stability of mixed
states does not exist in Model 3. Note here that
the analytical expression of the cross-talk noise
variance expressed by eq.\ (\ref{equation.model1r}) or eq.\ (\ref{equation.model2r})
is explicitly expressed as a function of susceptibility $U$, which is a
kind of a cross-talk noise-enhancement factor
caused by the full feedback nature of the model. 
However, the analytical expression of the cross-talk noise
variance for Model 3, $ \sigma $ in eq.\ (\ref{equation.Model3r}), is
{\it not} dependent on the susceptibility $U$. 
The above results show that the bi-stability is due to
not only the hierarchy of stored patterns but also analytical
expression of the cross-talk noise variance,
which is  a function of susceptibility for the model. 
Naturally, the bi-stability may exist in more complex hierarchies because the corresponding
cross-talk noise variance is explicitly expressed by a function of
susceptibility as shown in eq.\ (\ref{equation.xigr}).
From these results, we also found that the bi-stability of mixed states
does not exist in a strongly diluted system \cite{Derrida1987}.

Finally, we discuss the relation between findings concerning the retrieval process in Fig. 4
and the physiological experiments of face-responsive neurons in the IT cortex by 
Sugase et al \cite{Sugase1999}. 
Recently, Sugase et al.\ have analyzed the
time-course of information carried by the firing of face-responsive
neurons in the IT cortex, while performing a fixation task of monkey and 
human faces with various expressions, and simple geometrical shapes.
They found that the initial transient firing correlated well with a
rough categorization (e.g., face vs. non-face stimuli). Their results
suggest that the neuron firing pattern is initially a superposition of
patterns representing different faces or expressions, but it then
converges to a single pattern representing a specific face or expression.
We found that the retrieval dynamics of Model 1, shown in Fig.\ \ref{dynamics}, can qualitatively replicate the
temporal dynamics of face-responsive neurons as follows \cite{OkadaNIPS}. Initially, the 
network state approaches a mixed state ($\mbox{\boldmath $\tilde{ \eta}$}^{1}$ or $\mbox{\boldmath
$\eta$}^{1}$ ) that is a superposition of patterns representing
different persons or expressions. After that it diverges from the mixed state, and
finally converges to a single memory pattern ($\mbox{\boldmath
$\xi$}^{1,1}$ ) representing a specific person or expression as shown in
Fig.\ \ref{dynamics}. From the above results, we expect that the present system may mimic the temporal
dynamics of the face-responsive neurons \cite{OkadaNIPS}. Details will
be discussed elsewhere.

\acknowledgments This work was supported in part 
by Grants-in-Aids \#09308010 and \#11145229 for Scientific Research
from the Ministry of Education, Science, Sports and Culture of Japan.
We are indebted to Tomoki Fukai for useful discussion.
\appendix
\section{}
The internal potential of the $i$-the neuron $h_i$ in the equilibrium
state of eq.\ (\ref{dynamics_gene.rule}) is
\begin{eqnarray}
h_i &=& \sum_{j \neq i}^{N} J_{ij} x_j ,\\
     &=& \sum_{\rho=1}^{\hat{\alpha}N} \kappa_{\rho}
\bar{\sigma}^{\rho}_i  \bar{m}^{\rho}
-
\frac{1}{N} \sum_{\rho=1}^{\hat{\alpha} N} \kappa_{\rho} x_i. \label{eqA1}
\end{eqnarray}
The output $x_i$ can be formally expressed as
\begin{eqnarray}
x_i &=& F( \sum_{\rho=1}^{\hat{\alpha} N} \kappa_{\rho}
\bar{\sigma}^{\rho}_i  \bar{m}^{\rho}
-
\frac{1}{N} \sum_{\rho=1}^{\hat{\alpha} N} \kappa_{\rho} x_i ),\\
&=& \tilde{F}( \sum_{\rho=1}^{\hat{\alpha} N} \kappa_{\rho}
\bar{\sigma}^{\rho}_i  \bar{m}^{\rho}),
\end{eqnarray}
where the function $\tilde{F}(\cdot)$ will be determined later as shown
in eq.\ (26). Here, we assume
that the equilibrium state $\mbox{\boldmath $x$}$ has nonzero overlaps
with $\bar{s}$ rotated patterns $\bar{\mbox{\boldmath $\sigma$}}^{\rho} \ (\rho=1
,2,...,\bar{s})$. The residual overlap $\bar{m}^{\rho} \sim O(1/\sqrt{N}),
\ (s+1 \le \rho \le \hat{\alpha} N)$ can be derived by using Taylor expansion
\begin{eqnarray}
\bar{m}^{\rho} &=& \frac{1}{N} \sum_{i=1}^{N} \bar{\sigma}^{\rho}_i
\tilde{F}( \sum_{\rho'=1}^{\hat{\alpha}N} \kappa_{\rho'}
\bar{\sigma}^{\rho'}_i  \bar{m}^{\rho'}), \\
                 &=& \frac{1}{N} \sum_{i=1}^{N} \bar{\sigma}^{\rho}_i x_i^{(\rho)}
+ \kappa_{\rho} U \bar{m}^{\rho}, \\
                 &=& \frac{1}{N(1-\kappa_{\rho} U )}
\sum_{i=1}^{N} \bar{\sigma}^{\rho}_i x_i^{(\rho)} \label{eqA2}
\end{eqnarray}
where
\begin{eqnarray}
x_i^{(\rho)} &=& \tilde{F}( \sum_{\rho' \neq \rho}^{\hat{\alpha}N} \kappa_{\rho'}\bar{\sigma}^{\rho'}_i
\bar{m}^{\rho'}) ,\\
{x_i^{(\rho)}}' &=& \tilde{F}'(\sum_{\rho' \neq \rho}^{\hat{\alpha}N} \kappa_{\rho'}\bar{\sigma}^{\rho'}_i
\bar{m}^{\rho'}) ,\\
U &=& \frac{1}{N} \sum_{i=1}^N {x'}_i^{(\rho)}.
\end{eqnarray}
Substituting eq.\ (\ref{eqA2}) into eq.\ (\ref{eqA1}), we obtain
\begin{eqnarray}
h_i &=& \sum_{\rho=1}^{\bar{s}} \kappa_{\rho}
\bar{\sigma}^{\rho}_i  \bar{m}^{\rho}
+ \Gamma x_i + \bar{z}_i,
\end{eqnarray}
where $\Gamma$ is defined as
\begin{eqnarray}
\Gamma &=& \frac{1}{N} \sum_{\rho=\bar{s}+1}^{\hat{\alpha}N} \frac{{\kappa_{\rho}}^2 U}
{1-\kappa_{\rho}U},
\end{eqnarray}
and $\bar{z}_i$ is the effective noise so that,
\begin{eqnarray}
\bar{z}_i &=& \frac{1}{N} \sum_{\rho =\bar{s}+1}^{\hat{\alpha}N}\sum_{j\neq
 i}^{N}\frac{\kappa_{\rho}}
{1-\kappa_{\rho}U}\bar{\sigma}_i^{\rho}\bar{\sigma}_j^{\rho}x_j^{(\rho)}.
\end{eqnarray}
Note that $\bar{z}_i$ is a summation of uncondensed patterns with
$<\bar{z}>=0$ and $<\bar{z}^2>=\hat{\alpha}r$,
\begin{eqnarray}
r &=& \frac{1}{\hat{\alpha}N} \sum_{\rho =\bar{s}+1}^{\hat{\alpha}N}
\frac{\kappa_{\rho}^2}{(1-\kappa_{\rho}U)^2}
\frac{1}{N}\sum_{j=1}^{N} (x_j^{(\rho)})^2.
\end{eqnarray}
Let us express the sum of $\kappa_{\rho}$
in terms of an integration along continuous eigenvalue $\kappa
( \frac{\rho}{\hat{\alpha} N} ) \equiv \kappa_{\rho}$ for $p, N
\rightarrow \infty$,
\begin{eqnarray}
r &=& q \int_{0}^{1} du \frac{{\kappa(u)}^2
 }{(1-\kappa(u)U)^2} ,\\
q &=& \frac{1}{N} \sum_{i=1}^N (x_i)^2 ,\\
\Gamma &=& \hat {\alpha} \int_{0}^{1} du \frac{{\kappa(u)}^2 U}{1-\kappa(u)U}.
\end{eqnarray}
We can obtain eqs.\ (\ref{equation.gm}) - (\ref{equation.gGamma}) by
replacing $x_i$ with $Y$.
\section{}
Let us transform the order parameter equations for the mixed state in
Model 1 into the equation in one variable. We introduce $y^{\nu}$ in the
following equation,
\begin{equation}
y^{\nu}=\frac{m^{1,\nu}}{\sqrt{2 \alpha r}}.
\end{equation}
Since $y^{\nu}=y$ $(\nu=1,2,3,...,s)$ holds in the mixed state, we can
express the order parameter equations for the mixed state by the following
equation in $y$,
\begin{eqnarray}
\Phi(y,  \alpha,  b,  s) &=& 0 ,\\
\Phi(y,  \alpha,  b,  s) & \equiv &  \sum_{\nu=1}^s \biggl( \frac{ \lambda_{\nu}^{2}
 }{ ( w(y) - \lambda_{\nu}\theta(y) )^2} \biggl) -1 ,\\
\theta(y) &\equiv & \sqrt{ \frac{2}{\pi \alpha } }
< \exp (-(\sum_{ \nu}^{s} y \xi^{1,   \nu})^2 )  >_{{\mbox{\boldmath $\xi$}}^1}, \\
w(y) & \equiv & \frac{1}{\sqrt{2 \alpha } y} <
\xi^{1,  \nu}\mbox{erf}(\sum_{ \nu}^{s} y \xi^{1,   \nu})
 >_{{\mbox{\boldmath $\xi$}}^1} ,
\end{eqnarray}
where  $\lambda_{\nu}$ is the $\nu$-th eigenvalue of the matrix $\bf B$.
The order parameter equations for the mixed states in Model 2 can be also
transformed into the equation in one variable as follows,
\begin{eqnarray}
\Psi(y,  \alpha, b,  s) &=& \sqrt{2 \alpha}y \Big(  
 \sqrt{ \frac{2}{\pi \alpha } }
< \exp (-(\sum_{ \nu}^{s} y \xi^{1,   \nu})^2 )  >_{{\mbox{\boldmath
$\xi$}}^1} - 
<
\xi^{1,  \nu}\mbox{erf}(\sum_{ \nu}^{s} y \xi^{1,   \nu})
 >_{{\mbox{\boldmath $\xi$}}^1} \nonumber \\
&=&0 .
\end{eqnarray}



\begin{figure}
 \caption{Variation of critical loading rate of memory pattern
or mixed states with $b$. Two kinds of mixed states coexist in triangular region. }
 \label{phasebalpha}
\end{figure}

\begin{figure}
 \caption{(a) Typical example ($b=0.61, s=3$) of $\alpha$ dependency of overlap between equilibrium
 state and retrieval pattern. (b) Solutions of transformed
order parameter equation in one variable around bi-stable region when $\alpha^{\eta}_c 
 < \alpha^{\tilde{\eta}}_c$}
\label{b61malpha}  
\end{figure}

\begin{figure}
\caption{(a) Typical example ($b=0.55, s=3$) of $\alpha$ dependency of overlap between equilibrium
state and retrieval pattern. (b) Solutions of transformed
order parameter equation in one variable around bistable region when
 $\alpha^{\tilde{\eta}}_c  < \alpha^{\eta}_c$ }
\label{b55malpha}
\end{figure}

\begin{figure}
\caption{Retrieval processes from various initial states in computer
 simulation when $b=0.475$, $s=3$, $\alpha=0.0087$, $N=40000$. }
\label{dynamics}
\end{figure}

\end{document}